\documentstyle[psfig]{mn}

\newif\ifAMStwofonts

\ifoldfss
  \ifCUPmtlplainloaded \else
    \NewTextAlphabet{textbfit} {cmbxti10} {}
    \NewTextAlphabet{textbfss} {cmssbx10} {}
    \NewMathAlphabet{mathbfit} {cmbxti10} {} 
    \NewMathAlphabet{mathbfss} {cmssbx10} {} 
  \fi
  \ifAMStwofonts
    \ifCUPmtlplainloaded \else
      \NewSymbolFont{upmath} {eurm10}
      \NewSymbolFont{AMSa} {msam10}
      \NewMathSymbol{\upi}     {0}{upmath}{19}
      \NewMathSymbol{\umu}     {0}{upmath}{16}
      \NewMathSymbol{\upartial}{0}{upmath}{40}
      \NewMathSymbol{\leqslant}{3}{AMSa}{36}
      \NewMathSymbol{\geqslant}{3}{AMSa}{3E}

       \let\le=\leqslant
       \let\ge=\geqslant
    \fi
  \fi
\fi 

\ifnfssone
  \newmathalphabet{\mathit}
  \addtoversion{normal}{\mathit}{cmr}{m}{it}
  \addtoversion{bold}{\mathit}{cmr}{bx}{it}
  \def\textbfit{\protect\txtbfit}
  
  \long\def\txtbfit#1{{\fontfamily{cmr}\fontseries{bx}\fontshape{it}%
    \selectfont #1}}

  \newmathalphabet{\mathbfit} 
  \addtoversion{normal}{\mathbfit}{cmr}{bx}{it}
  \addtoversion{bold}{\mathbfit}{cmr}{bx}{it}
  \newmathalphabet{\mathbfss} 
  \addtoversion{normal}{\mathbfss}{cmss}{bx}{n}
  \addtoversion{bold}{\mathbfss}{cmss}{bx}{n}
  \ifAMStwofonts
    \ifCUPmtlplainloaded \else
      \UseAMStwoboldmath
      \makeatletter
      \new@mathgroup\upmath@group
      \define@mathgroup\mv@normal\upmath@group{eur}{m}{n}
      \define@mathgroup\mv@bold\upmath@group{eur}{b}{n}
      \edef\UPM{\hexnumber\upmath@group}
      \new@mathgroup\amsa@group
      \define@mathgroup\mv@normal\amsa@group{msa}{m}{n}
      \define@mathgroup\mv@bold\amsa@group{msa}{m}{n}
      \edef\AMSa{\hexnumber\amsa@group}
      \makeatother
      \mathchardef\upi="0\UPM19
      \mathchardef\umu="0\UPM16
      \mathchardef\upartial="0\UPM40
      \mathchardef\leqslant="3\AMSa36
      \mathchardef\geqslant="3\AMSa3E

       \let\le=\leqslant
       \let\ge=\geqslant
    \fi
  \fi
\fi 

\ifnfsstwo
  \def\textbfit{\protect\txtbfit}
  
  \long\def\txtbfit#1{{\fontfamily{cmr}\fontseries{bx}\fontshape{it}%
    \selectfont #1}}

  \DeclareMathAlphabet{\mathbfit}{OT1}{cmr}{bx}{it}
  \SetMathAlphabet\mathbfit{bold}{OT1}{cmr}{bx}{it}
  \DeclareMathAlphabet{\mathbfss}{OT1}{cmss}{bx}{n}
  \SetMathAlphabet\mathbfss{bold}{OT1}{cmss}{bx}{n}
  \ifAMStwofonts
    \ifCUPmtlplainloaded \else
      \DeclareSymbolFont{UPM}{U}{eur}{m}{n}
      \SetSymbolFont{UPM}{bold}{U}{eur}{b}{n}
      \DeclareSymbolFont{AMSa}{U}{msa}{m}{n}
      \DeclareMathSymbol{\upi}{0}{UPM}{"19}
      \DeclareMathSymbol{\umu}{0}{UPM}{"16}
      \DeclareMathSymbol{\upartial}{0}{UPM}{"40}
      \DeclareMathSymbol{\leqslant}{3}{AMSa}{"36}
      \DeclareMathSymbol{\geqslant}{3}{AMSa}{"3E}

       \let\le=\leqslant
       \let\ge=\geqslant
    \fi
  \fi
\fi 

\ifCUPmtlplainloaded \else
  \ifAMStwofonts \else 
    \def\upi{\pi}
    \def\umu{\mu}
    \def\upartial{\partial}
  \fi
\fi

\title[Light element evolution resulting from {\it WMAP\/}]
      {Light element evolution resulting from {\textbfit{WMAP\/}} data}
\author[D. Romano et al.]
       {Donatella Romano,$^1$
       Monica Tosi,$^1$ 
       Francesca Matteucci,$^2$ 
       and Cristina Chiappini$^3$
       \thanks{E-mail: romano@bo.astro.it (DR); tosi@bo.astro.it (MT); 
       matteucci@ts.astro.it (FM); chiappini@ts.astro.it (CC)}\\
       $^1$INAF, Osservatorio Astronomico di Bologna,
           Via Ranzani 1, I-40127 Bologna, Italy\\
       $^2$Dipartimento di Astronomia, Universit\`a di Trieste,
           Via G.B. Tiepolo 11, I-34131 Trieste, Italy\\
       $^3$INAF, Osservatorio Astronomico di Trieste,
           Via G.B. Tiepolo 11, I-34131 Trieste, Italy
       }
\date{Accepted .
      Received ;
      in original form }

\pagerange{\pageref{firstpage}--\pageref{lastpage}}
\pubyear{2003}

\begin{document}

\maketitle

\label{firstpage}

   \begin{abstract}
   The recent determination of the baryon-to-photon ratio from {\it WMAP\/} 
   data by Spergel et al. (2003) allows one to fix with unprecedented 
   precision the primordial abundances of the light elements D, $^3$He, $^4$He 
   and $^7$Li in the framework of the standard model of big bang 
   nucleosynthesis. We adopt these primordial abundances and discuss the 
   implications for Galactic chemical evolution, stellar evolution and 
   nucleosynthesis of the light elements. The model predictions on D, $^3$He 
   and $^4$He are in excellent agreement with the available data, while a 
   significant depletion of $^7$Li in low-metallicity stars is required to 
   reproduce the \emph{Spite plateau}.
   \end{abstract}


   \section{Introduction}

   The standard theory of big bang nucleosynthesis (SBBN; e.g., Boesgaard \& 
   Steigman 1985; Steigman 1989; Walker et al. 1991) accurately predicts the 
   primordial abundances of the light elements D, $^3$He, $^4$He and $^7$Li, 
   as a function of the cosmic baryon density, $\rho_{\mathrm{b}} \propto 
   \Omega_{\mathrm{b}} \, h^2$, or, equivalently, of the baryon-to-photon 
   ratio, $\eta$. Since the baryon density is the sole parameter in the SBBN, 
   observations of D, $^3$He, $^4$He and $^7$Li in astrophysical environments 
   not yet affected by subsequent stellar evolution offer a direct way to 
   infer the baryon density of the universe and to assess whether the SBBN 
   theory correctly describes the first three minutes of the hot early 
   universe. Nowadays, there is still disagreement on which is the value of 
   $\eta$ as inferred from {\emph{i)}} limits on the primordial deuterium 
   abundance from high-redshift absorption systems (mostly Damped Lyman 
   $\alpha$ or Lyman limit absorbers; e.g., Carswell et al. 1994; Songaila et 
   al. 1994; Tytler, Fan \& Burles 1996; Rugers \& Hogan 1996; Songaila, 
   Wampler \& Cowie 1997; Webb et al. 1997; Burles \& Tytler 1998; Levshakov, 
   Kegel \& Takahara 1998; O'Meara et al. 2001; Pettini \& Bowen 2001; 
   Levshakov et al. 2002); {\emph{ii)}} estimates of the primordial $^4$He 
   abundance from extragalactic H\,{\small II} regions (e.g., Olive, Skillman 
   \& Steigman 1997; Izotov et al. 1999; Peimbert, Peimbert \& Luridiana 2002; 
   see also Pagel 2000 for a review) and {\emph{iii)}} estimates of the 
   pristine $^7$Li content in old, metal-poor ([Fe/H] $\le -1.5$ dex), warm 
   ($T_{\mathrm{eff}} \ge 5700$ K) dwarf stars in the solar neighbourhood 
   (e.g., Spite \& Spite 1982; Spite, Maillard \& Spite 1984; Spite \& Spite 
   1986; Rebolo, Molaro \& Beckman 1988; Thorburn 1992, 1994; Bonifacio \& 
   Molaro 1997; Vauclair \& Charbonnel 1995, 1998; Th\'eado \& Vauclair 2001; 
   Pinsonneault et al. 1999, 2002). Measurements of $^3$He/H in H\,{\small II} 
   regions in the outer parts of the Galactic disc should provide values very 
   close to the primordial one, owing to the slow evolution of the disc in 
   these regions, but they appear to poorly constrain the $\eta$ range (e.g., 
   Rood et al. 1998; Bania, Rood \& Balser 2002).

   Data on cosmic microwave background (CMB) anisotropies provide an 
   alternative, independent method for constraining $\eta$. The first release 
   of results from the {\it Wilkinson Microwave Anisotropy Probe\/} ({\it 
   WMAP\/}; Bennett et al. 2003; Spergel et al. 2003) makes the CMB the prime 
   cosmic baryometer, owing to the high {\it WMAP\/} precision. A combination 
   of {\it WMAP\/} data with other finer scale CMB experiments (ACBAR -- Kuo 
   et al. 2002 -- and CBI -- Pearson et al. 2002) and with astronomical 
   measurements of the power spectrum (2dF Galaxy Redshift Survey measurements 
   -- Percival et al. 2001 -- and Lyman $\alpha$ forest data -- Croft et al. 
   2002; Gnedin \& Hamilton 2002) gives $\Omega_{\mathrm{b}} \, h^2 = 0.0224 
   \pm 0.0009$, or, equivalently, $\eta_{\mathrm{10, \, CMB}} = 
   6.1^{+0.3}_{-0.2}$ (Spergel et al. 2003)\footnote{Notice that 
   $\eta_{\mathrm{10}, \, \mathit{WMAP}} = 6.5^{+0.4}_{-0.3}$ from {\it 
   WMAP\/} data only.}, where $\eta_{\mathrm{10}} \equiv 10^{10} \, \eta$. By 
   adopting the SBBN predictions and using this $\eta$ value, one obtains the 
   primordial abundances of D, $^3$He, $^4$He and $^7$Li, and can use them to 
   compare with observations in low-metallicity environments and to gain 
   insight into chemical evolution.

%
   \begin{figure}
   \centerline{\psfig{figure=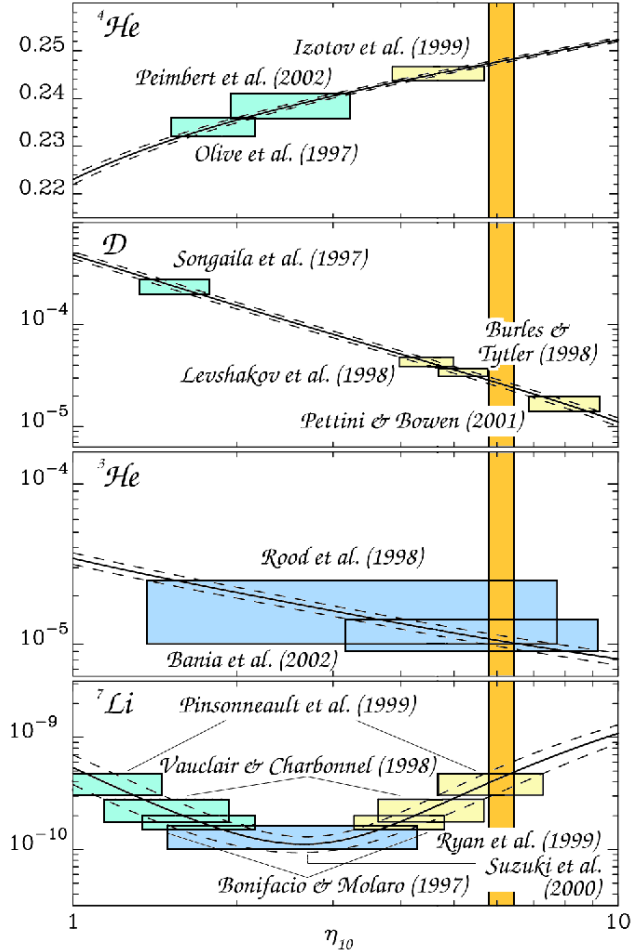,width=0.48\textwidth,angle=0}}
      \caption{Solid lines: primordial abundances of $^4$He (mass fraction), 
	      D, $^3$He, $^7$Li (number ratios relative to hydrogen) as a 
	      function of $\eta_{\mathrm{10}} \equiv 10^{10} \, \eta$ from 
	      SBBN. The dashed lines are 1$\sigma$ deviations. Theoretical 
	      predictions are from Fiorentini et al. (1998). The data quoted 
	      in the Introduction are shown with 1$\sigma$ boxes. The vertical 
	      band represents the {\it WMAP\/} $\eta_{\mathrm{10}}$ range.
              }
         \label{FigLightPrim}
   \end{figure}
%

   The aim of this paper is to analyse the effect of adopting primordial 
   abundances predicted by SBBN constrained by {\it WMAP\/} on the chemical 
   evolution of the Milky Way and to gain insights into the mechanisms of 
   light element production/destruction inside stars.


   \section{Primordial abundances}

   In Fig.~\ref{FigLightPrim} we show the primordial abundances of D, $^3$He, 
   $^4$He and $^7$Li predicted by the SBBN model as a function of 
   $\eta_{\mathrm{10}}$, with their 1$\sigma$ deviations. Theoretical 
   predictions are from Fiorentini et al. (1998). The values inferred from the 
   observations quoted in the Introduction are also displayed with 1$\sigma$ 
   boxes.

   The primordial abundance by mass of $^4$He ($Y_{\mathrm{p}}$) is derived by 
   analysing the helium content in low-metallicity extragalactic H\,{\small 
   II} regions. Since even in the lowest metallicity H\,{\small II} regions 
   some $^4$He must be produced by stars along with the heavy elements, it is 
   necessary to correlate $Y_{\mathrm{obs}}$ with metallicity and to 
   extrapolate to zero metallicity in order to infer $Y_{\mathrm{p}}$ 
   (Peimbert \& Torres-Peimbert 1974). Olive et al. (1997) found 
   $Y_{\mathrm{p}} = 0.234 \pm 0.002$ for their full data set and an even 
   lower value, $Y_{\mathrm{p}} = 0.230 \pm 0.003$, for a subset of H\,{\small 
   II} regions with the lowest metallicity. Izotov et al. (1999) used 
   high-quality spectroscopic observations of the two most metal-deficient 
   blue compact galaxies known, I\,Zw\,18 and SBS\,0335-052, to determine 
   $Y_{\mathrm{p}}$. The weighted mean helium mass fraction in the two 
   galaxies, $Y = 0.2462 \pm 0.0015$, after correction for the stellar $^4$He 
   production implies a primordial $^4$He mass fraction $Y_{\mathrm{p}} = 
   0.2452 \pm 0.0015$. This value corresponds to $\eta_{\mathrm{10}} = 
   4.7^{+1.0}_{-0.8}$, while Olive et al.'s value translates into 
   $\eta_{\mathrm{10}} = 1.8 \pm 0.3$. This discrepancy could be less severe, 
   if corrections for unseen neutral hydrogen and/or helium apply to Izotov et 
   al.'s estimate of $Y_{\mathrm{p}}$ (Gruenwald, Steigman \& Viegas 2002). 
   Very recently, Peimbert et al. (2002) have obtained $Y_{\mathrm{p}} = 
   0.2384 \pm 0.0025$ from the helium abundances of five objects including the 
   three low-metallicity objects with the best line determinations and the two 
   objects with the lowest metallicity. They used the temperature derived from 
   the He\,{\small I} lines, which is always smaller than that derived from 
   the [O\,{\small III}] lines, and considered the collisional contribution to 
   the Balmer-line intensities, never taken into account in the other recent 
   $Y_{\mathrm{p}}$ determinations in the literature. Indeed, the temperature 
   structure, the collisional excitation of the hydrogen lines, and the 
   ionization structure are problems affecting the $Y_{\mathrm{p}}$ 
   determination that need to be further analysed. The determination of 
   $Y_{\mathrm{p}}$ by Peimbert et al. favors $\eta$ values in the range 
   $(2-3) \times 10^{-10}$.

   An alternative approach to derive $Y_{\mathrm{p}}$ from Galactic globular 
   clusters was suggested by Iben (1968) and Iben \& Faulkner (1968). This 
   method originally provided low values of $Y_{\mathrm{p}} = 0.23 \pm 0.02$ 
   (Buzzoni et al. 1983), but was affected by significant uncertainties both 
   on the star counts and on the stellar evolution theory. Very recent 
   application of the method to much more complete data sets and with updated 
   stellar models (Cassisi, Salaris \& Irwin 2003) indicates instead 
   $Y_{\mathrm{p}} \simeq 0.243-0.244$.

   Observations of D in high-redshift QSO absorbers provide an estimate of the 
   primordial deuterium abundance. The controversy on whether the primordial 
   abundance of deuterium was a few times $10^{-5}$ or one order of magnitude 
   higher (e.g., Carswell et al. 1994; Songaila et al. 1994; Tytler et al. 
   1996; Rugers \& Hogan 1996; Songaila et al. 1997; Webb et al. 1997; Burles 
   \& Tytler 1998) seems now largely solved: several low D/H values have been 
   reported in the last few years (e.g., Levshakov et al. 1998; O'Meara et al. 
   2001; Pettini \& Bowen 2001; Levshakov et al. 2002 -- Table 1) and it has 
   been argued that D/H values higher than a few $10^{-5}$ are likely to be 
   due to contamination by H\,{\small I} interlopers at velocities similar to 
   the isotope shift (e.g., Kirkman et al. 2001).
%
   \begin{table*}
      \caption[]{D/H measurements at high redshift.}
     $$ 
         \begin{array}{p{0.2\linewidth}l p{0.6\linewidth}l}
            \hline
            \noalign{\smallskip}
            Authors     & 10^5 \, \mathrm{D/H} &  Notes \\
            \noalign{\smallskip}
            \hline
            \noalign{\smallskip}
Levshakov et al. (1998) & 4.1-4.6       & Range common to different analyses \\
O'Meara et al. (2001)   & 3.0 \pm 0.4   & Best value from measurements in 
                                          different objects \\
                        & 2.54 \pm 0.23 & Lyman limit system at 
					  z$_{\mathrm{abs}}$ = 2.536 
                                          along the line of sight of 
					  HS\,0105+1619 \\
Pettini \& Bowen (2001) & 1.65 \pm 0.35 & DLA at z$_{\mathrm{abs}}$ = 2.0762 
                                          along the line of sight of 
					  Q\,2206-199 \\
Levshakov et al. (2002) & 3.75 \pm 0.25 & DLA at z$_{\mathrm{abs}}$ = 3.025 
                                          along the line of sight of 
					  Q\,0347-3819 \\
            \noalign{\smallskip}
            \hline
         \end{array}
     $$ 
   \end{table*}
%

   $^3$He is not a very good baryometer. Rood et al. (1998) suggested 
   ($^3$He/H)$_{\mathrm{p}} = 1.5^{+1.0}_{-0.5} \times 10^{-5}$ as a 
   reasonable estimate for the primordial $^3$He. A smaller range, 
   ($^3$He/H)$_{\mathrm{p}} = (1.1 \pm 0.2) \times 10^{-5}$, has been more 
   recently obtained by Bania et al. (2002) for a Galactic H\,{\small II} 
   region located at $R_{\mathrm{G}} = 16.9$ kpc, namely, for a region where 
   stellar activity should not have contaminated the pristine $^3$He abundance 
   significantly. As can be seen in Fig.~\ref{FigLightPrim}, even this latter, 
   more precise determination of ($^3$He/H)$_{\mathrm{p}}$ does not 
   significantly constrain the value of $\eta_{\mathrm{10}}$, because of the 
   weak dependence of ($^3$He/H)$_{\mathrm{p}}$ on the baryon density. 
   However, it points against the low $\eta_{10}$ range ($\eta_{10} < 3$) 
   suggested by the high-D, low-$^4$He determinations.

   The actual value of the primordial $^7$Li abundance is still controversial, 
   with some people favoring the hypothesis that the so-called {\emph{Spite 
   plateau}} is truly representative of the pristine $^7$Li content of halo 
   dwarfs, and hence of the primordial $^7$Li abundance (Spite \& Spite 1982; 
   Spite et al. 1984, Spite \& Spite 1986; Rebolo et al. 1988; Spite et al. 
   1996; Bonifacio \& Molaro 1997), and other authors suggesting the existence 
   of non-standard depletion processes to reproduce the observed plateau value 
   starting from a higher primordial one (e.g., Thorburn \& Beers 1993; 
   Vauclair \& Charbonnel 1995, 1998; Th\'eado \& Vauclair 2001; Pinsonneault 
   et al. 1999, 2002). On the other hand, the primordial lithium abundance 
   could be even lower than that shared by warm halo dwarfs because of 
   significant early contribution from Galactic chemical evolution ($\alpha + 
   \alpha$ fusion mechanism and stellar nucleosynthesis; e.g., Ryan, Norris \& 
   Beers 1999; Suzuki, Yoshii \& Beers 2000). A set of representative 
   primordial abundances of $^7$Li is shown in Fig.~\ref{FigLightPrim}. Notice 
   that the most recent estimate of the plateau value from Bonifacio (2002), 
   $\log \varepsilon$($^7$Li)$_{\mathrm{p}}$\footnote{$\log 
   \varepsilon$($^7$Li) = $\log$(N$_{\mathrm{^7Li}}$/N$_{\mathrm{H}}$) + 12, 
   where N$_{\mathrm{^7Li}}$ and N$_{\mathrm{H}}$ are the abundances by number 
   of $^7$Li and H, respectively.} = 2.317 $\pm$ 0.014$_{1\sigma}$ $\pm$ 
   0.05$_{\mathrm{sys}}$, is not shown, because it refers to a preliminary 
   data analysis published inso far only as conference proceedings. The quoted 
   value results from the analysis of a high-quality sample of 22 stars 
   satisfying the conditions ${T}_{\mathrm{eff}} >$ 5700 K, [Fe/H] $\le -$1.5, 
   taking into account the effects of standard depletion and non-LTE. This 
   value is higher than previous estimates by Spite \& Spite (1982) [$\log 
   \varepsilon$($^7$Li)$_{\mathrm{p}} = 2.05 \pm 0.15$] and Bonifacio \& 
   Molaro (1997) [$\log \varepsilon$($^7$Li)$_{\mathrm{p}}$ = 2.238 $\pm$ 
   0.012$_{1\sigma}$ $\pm$ 0.05$_{\mathrm{sys}}$], but always lower than those 
   suggested if non-standard stellar depletion is acting in low-mass, warm 
   halo stars.

   The first release of results from {\it WMAP\/} allows us to independently 
   constrain the range of values of $\eta_{\mathrm{10}}$ with unprecedented 
   precision. A value of $\eta_{\mathrm{10}} = 6.1^{+0.3}_{-0.2}$ is found 
   (Spergel et al. 2003). The $\eta_{\mathrm{10}}$ range is shown in 
   Fig.~\ref{FigLightPrim} as a vertical band. It is clearly seen that the 
   low-$^4$He and high-D values are completely ruled out. In particular, for D 
   this was already a prediction from Galactic chemical evolution (GCE) models 
   (Steigman \& Tosi 1992; Galli et al. 1995; Prantzos 1996; Tosi et al. 1998; 
   Chiappini, Renda \& Matteucci 2002). In fact, the level of stellar 
   astration required by chemical evolution models for deuterium to match its 
   present interstellar medium (ISM) abundance starting from a high primordial 
   value is inconsistent with a number of observational constraints for the 
   Milky Way.

   In the following, we adopt the primordial abundances resulting from the 
   $\eta_{\mathrm{10}}$ range derived from {\it WMAP\/} data and study the 
   evolution of the light elements D, $^3$He, $^4$He and $^7$Li in the Galaxy. 
   We discuss the implications of these model results on stellar evolution and 
   nucleosynthesis.


   \section{Light element evolution in the Milky Way}

   We adopt two different chemical evolution models for the Milky Way, the one 
   developed by Chiappini et al. (1997, 2001) and that developed by Tosi 
   (1988a,\,b -- model called Tosi 1). Both models satisfy the main 
   observational constraints for the solar neighbourhood as well as the whole 
   Galactic disc. In the following, we underline the major differences and 
   similarities between the models.

   In both models, the Galactic disc is divided into concentric rings without 
   radial flows between them and the disc of the Galaxy is built up 
   \emph{inside-out} (Matteucci \& Fran\c cois 1989) from gas infalling from 
   outside (see also Larson 1976; Tinsley \& Larson 1978; Tosi 1982). In 
   Chiappini et al.'s model, the infall rate is exponentially decreasing in 
   time, with an $e$-folding time increasing with increasing Galactic radius. 
   In particular, $\tau_{\mathrm{D}}(R_{\mathrm{G}}) = 1.03 \times 
   R_{\mathrm{G}} - 1.27$ Gyr (Romano et al. 2000). This leads to a faster 
   disc formation in the innermost regions. The infalling gas has a primordial 
   chemical composition. In the model of Tosi, a constant (in time) and 
   uniform (in space) infall density of 0.004 $M_\odot$ kpc$^{-2}$ yr$^{-1}$ 
   is assumed, so that more mass is accreted in the outer disc than in the 
   inner one; the infall law is formally exponential and decreasing in time, 
   but practically constant. The infalling gas is slightly metal-enriched, 
   $Z_{\mathrm{inf}}$ = 0.2 $Z_\odot$ (see Tosi 1988b). In Chiappini et al.'s 
   model, the star formation rate at a given radius $R_{\mathrm{G}}$ and time 
   $t$ is explicitly dependent on $\Sigma_{\mathrm{gas}}(R_{\mathrm{G}}, t)$, 
   $\Sigma_{\mathrm{tot}}(R_{\mathrm{G}}, t)$ and 
   $\Sigma_{\mathrm{tot}}(R_\odot, t_{\mathrm{Gal}})$, i.e., the gas and total 
   mass densities at that radius and time and the present-day total mass 
   density at the solar position. In Tosi's model the star formation is 
   explicitly dependent on the gas and total mass density currently observed 
   at each Galactocentric distance and exponentially decreasing in time (with 
   $e$-folding time 15 Gyr). In the model of Chiappini et al., a threshold in 
   the star formation process is considered, which makes the star formation in 
   the disc go to zero when the gas density falls below a critical value 
   ($\Sigma^{\mathrm{th}} = 7$ $M_\odot$ pc$^{-2}$). This threshold is not 
   considered in Tosi's model. The two models differ also in the adopted 
   initial mass functions (IMFs): Chiappini et al. adopt Scalo's (1986) IMF, 
   while Tosi adopts Tinsley's (1980) IMF. Both models avoid the instantaneous 
   recycling approximation, i.e., they account for the stellar lifetimes. 
   However, they adopt different stellar lifetimes.

%
   \begin{figure}
   \centerline{\psfig{figure=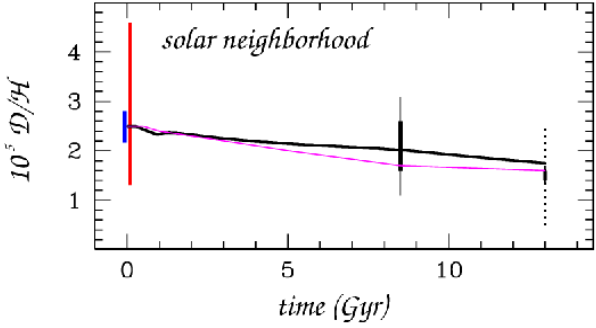,width=0.48\textwidth,angle=0}}
      \caption{Evolution of D/H in the solar neighbourhood. The thick 
	      continuous line are the predictions of Model CRM; the thin 
	      continuous line are the predictions of Model Tosi 1 (see text). 
	      The vertical bars at $t = 0$ represent the range of primordial D 
	      abundance from {\it WMAP\/} (smaller bar) and high-redshift QSO 
	      absorbers (bigger bar -- Levshakov et al. 1998; O'Meara et al. 
	      2001; Pettini \& Bowen 2001; Levshakov et al. 2002). The PSC 
	      data (Geiss \& Gloeckler 1998) and the local ISM value (Linsky 
	      1998) are also shown, at 1 and 2$\sigma$ (thick and thin bars at 
	      $t = 8.5$ and 13 Gyr, respectively). The vertical dotted line at 
	      $t = 13$ Gyr represents the range of abundances derived from 
	      Copernicus, HST-GHRS, IMAPS, STIS and FUSE data (see, e.g., 
	      Vidal-Madjar et al. 1998; Jenkins et al. 1999; Moos et al. 2002; 
	      Hoopes et al. 2003).
              }
         \label{FigDeutEv}
   \end{figure}
%

   For the stellar nucleosynthesis of D, $^3$He and $^4$He we assume that D is 
   only destroyed in stellar interiors\footnote{A mechanism for post big bang 
   deuterium production (interaction of protons accelerated in flares with the 
   stellar atmosphere creating a flux of free neutrons, which undergo 
   radiative capture on atmospheric protons resulting in D synthesis) has been 
   proposed (Mullan \& Linsky 1999), but we do not include it here. Indeed, it 
   has been recently proved that this production channel does not allow for D 
   production at a level which will reverse the monotonic decline of D 
   (Prodanovi\'c \& Fields 2003).}, while $^3$He and $^4$He are partly 
   produced and partly burnt to form heavier species (e.g., Dearborn, Steigman 
   \& Tosi 1996). The evolution of $^3$He is assumed to be strongly affected 
   by extra-mixing and subsequent Cool Bottom Processing (CBP) in low-mass 
   stars as prescribed by Sackmann \& Boothroyd (1999a,\,b). These 
   prescriptions are the same as those already adopted by Galli et al. (1997), 
   Palla et al. (2000) and Chiappini et al. (2002) to reproduce the observed 
   abundances of D and $^3$He. For $^7$Li, the adopted nucleosynthesis 
   prescriptions are those described by Romano et al. (1999, 2001). 

   \subsection{Evolution of D, $^{\mathbf{3}}$He and $^{\mathbf{4}}$He in the 
              Galaxy}

   Our models assume (D/H)$_{\mathrm{p}} = 2.5 \times 10^{-5}$, 
   ($^3$He/H)$_{\mathrm{p}} = 0.9 \times 10^{-5}$ and $Y_{\mathrm{p}} = 
   0.248$, consistently with the range of $\eta$ inferred from {\it WMAP\/} 
   data. These adopted values happen to be the same as adopted by Chiappini et 
   al. (2002) for their Model C-II, except for $^3$He, for which we assume a 
   slightly lower primordial abundance. Model C-II also assumes that 93 per 
   cent of the stars with mass $M <$ 2~$M_\odot$ suffer some extra-mixing, 
   resulting in an overall $^3$He destruction (see Chiappini et al. 2002 for 
   details; see also Galli et al. 1997; Charbonnel \& do Nascimento 1998). We 
   assume the same percentage throughout this paper as well, unless otherwise 
   stated.

%
   \begin{figure}
   \centerline{\psfig{figure=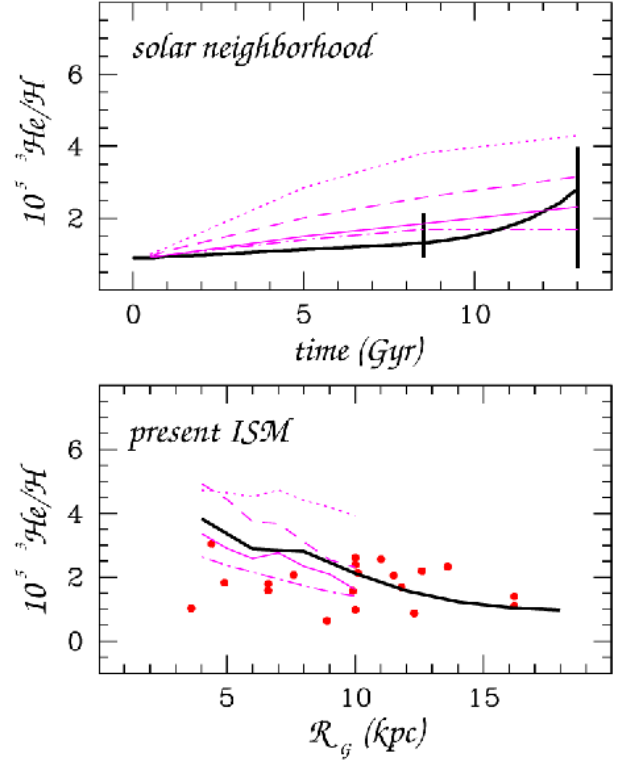,width=0.48\textwidth,angle=0}}
      \caption{Upper panel: evolution of $^3$He/H in the solar neighbourhood. 
	      Thick continuous line: Model CRM; thin continuous line: Model 
	      Tosi 1; dotted line: the same as Model Tosi 1, but assuming 
	      (D/H)$_{\mathrm{p}}$ = 20 $\times$ 10$^{-5}$; dashed line: the 
	      same as Model Tosi 1, but assuming (D/H)$_{\mathrm{p}}$ = 10 
	      $\times$ 10$^{-5}$; dot-dashed line: the same as Model Tosi 1, 
	      but with 100 per cent of low-mass stars suffering extra-mixing 
	      in conjunction with CBP occurrence. Data (vertical bars at $t = 
	      8.5$ and 13 Gyr) are from Geiss \& Gloeckler (1998). Lower 
	      panel: radial distribution of $^3$He/H for the same 
	      models in the upper panel. The dots are H\,{\small II} region 
	      data from Bania et al. (2002).
              }
         \label{Fig3HeEv}
   \end{figure}
%

   The adopted theoretical SBBN predictions are from Fiorentini et al. (1998). 
   The assumed ($^3$He/H)$_{\mathrm{p}}$ value is \emph{the lowest still 
   compatible with {\it WMAP\/} data} (at 1$\sigma$ level -- see 
   Fig.~\ref{FigLightPrim}). This choice is motivated by the fact that GCE 
   models always tend to overproduce $^3$He (especially if extra-mixing and 
   CBP are not included in the computations -- see, for instance, figure 4 of 
   Palla et al. 2000). Therefore, it is meaningful to start from the lowest 
   admissible value for ($^3$He/H)$_{\mathrm{p}}$. It is worth noticing that 
   different SBBN computations do not always lead to values of 
   (D/H)$_{\mathrm{p}}$, ($^3$He/H)$_{\mathrm{p}}$, $Y_{\mathrm{p}}$ and 
   ($^7$Li/H)$_{\mathrm{p}}$ which agree (within the errors) with those 
   adopted here. For instance, by adopting the SBBN predictions by Cyburt, 
   Fields \& Olive (2001), in light of the {\it WMAP\/} determination of 
   $\eta$ one gets (D/H)$_{\mathrm{p}} = 2.74^{+0.26}_{-0.16} \times 10^{-5}$, 
   ($^3$He/H)$_{\mathrm{p}} = 9.30^{+1.00}_{-0.67} \times 10^{-6}$, 
   $Y_{\mathrm{p}} = 0.2484^{+0.0004}_{-0.0005}$ and ($^7$Li/H)$_{\mathrm{p}} 
   = 3.76^{+1.03}_{-0.38} \times 10^{-10}$. Instead, for the OSU code, at 
   $\eta_{\mathrm{10}} = 6.14$, (D/H)$_{\mathrm{p}} = 2.57 \times 10^{-5}$ and 
   ($^7$Li/H)$_{\mathrm{p}} = 4.51 \times 10^{-10}$ (G. Steigman, private 
   communication). These results are fairly different from those listed above, 
   but similar to the ones from the Burles et al. (1999) code 
   [(D/H)$_{\mathrm{p}} = 2.60 \times 10^{-5}$ and ($^7$Li/H)$_{\mathrm{p}} = 
   4.91 \times 10^{-10}$]. However, we have checked that no appreciable 
   changes in the GCE predictions are produced by adopting these values rather 
   than those used here.

   In Fig.~\ref{FigDeutEv} we show the temporal behaviour of the deuterium 
   abundance in the solar neighbourhood predicted by Model CRM (the same as 
   Model C-II of Chiappini et al. 2002, but with a lower $^3$He primordial 
   abundance; thick continuous line) compared to that obtained by Model Tosi 1 
   (thin continuous line) starting from the same primordial deuterium 
   abundance and considering extra-mixing in 93 per cent of low-mass stars. It 
   can be seen that, notwithstanding the differences between the two models 
   mentioned above, the predicted behaviour is almost the same. In particular, 
   the two models predict nearly the same depletion factor for deuterium, 
   $X_{\mathrm{D, \, p}}/X_{\mathrm{D}, \, t_{\mathrm{Gal}}} \sim 1.5$, where 
   $X_{\mathrm{D, \, p}}$ and $X_{\mathrm{D}, \, t_{\mathrm{Gal}}}$ are the 
   primordial and present-day D abundances by mass, respectively. The vertical 
   bars at $t = 0$ represent the range allowed for the primordial deuterium 
   abundance from {\it WMAP\/} data (smaller bar) and from high-redshift QSO 
   absorbers (bigger bar -- Levshakov et al. 1998; O'Meara et al. 2001; 
   Pettini \& Bowen 2001; Levshakov et al. 2002). The vertical solid bars at 
   $t = 8.5$ Gyr are the Geiss \& Gloeckler (1998) Protosolar Cloud (PSC) data 
   at 1 and 2$\sigma$ (thick and thin lines, respectively), while the vertical 
   solid bars at $t = 13$ Gyr represent the local ISM value by Linsky (1998; 
   at 1 and 2$\sigma$ -- thick and thin solid lines, respectively). The dotted 
   line at $t = 13$ Gyr shows the most likely range of D/H variation as 
   allowed from measurements with Copernicus, HST-GHRS, IMAPS, STIS and FUSE 
   along several lines of sight (e.g., Vidal-Madjar et al. 1998; Jenkins et 
   al. 1999; Moos et al. 2002; Hoopes et al. 2003). However, notice that 
   values as high as D/H = 4 $\times$ 10$^{-5}$ have also been reported from 
   IUE observations (see Vidal-Madjar et al. 1998 for references, but see also 
   Vidal-Madjar \& Ferlet 2002). Starting from the {\it WMAP\/} primordial D 
   determination, the protosolar and local deuterium abundances are very well 
   reproduced by assuming that D is only destroyed in stars.

   In Fig.~\ref{Fig3HeEv}, upper panel, the evolution of $^3$He/H in the solar 
   neighbourhood predicted by the same models of Fig.~\ref{FigDeutEv} is 
   displayed, together with predictions from other three models, which differ 
   from Model Tosi 1 either in the assumed primordial deuterium abundance or 
   in the percentage of low-mass stars suffering extra-mixing. In particular, 
   the cases (D/H)$_{\mathrm{p}} = 20 \times 10^{-5}$, (D/H)$_{\mathrm{p}} = 
   10 \times 10^{-5}$ and 100 per cent of low-mass stars destroying $^3$He 
   through CBP are shown, as a dotted, dashed and dot-dashed thin line, 
   respectively.  The protosolar and local $^3$He abundances are from Geiss \& 
   Gloeckler (1998). It is seen that models starting from a high primordial 
   deuterium abundance are not able to reproduce the protosolar and/or local 
   $^3$He abundance, even allowing for $\sim 90$ per cent of extra-mixing in 
   low-mass stars. This gives further support to the argument that the 
   primordial D abundance cannot be higher than (D/H)$_{\mathrm{p}} \sim 4 
   \times 10^{-5}$ (Chiappini et al. 2002; see also Tosi et al. 1998). A model 
   where 100 per cent of low-mass stars destroy their $^3$He through CBP 
   results in an almost flat $^3$He behaviour from the time of the protosolar 
   nebula formation to now. This is clearly an extreme case, since the high 
   $^3$He abundances derived for a small sample of Galactic planetary nebulae 
   (e.g., Balser et al. 1997; Balser, Rood \& Bania 1999; Palla et al. 2000) 
   suggest that at least a minor fraction of the stars of 1\,--\,2 $M_\odot$ 
   do indeed produce $^3$He in significant quantities. In Fig.~\ref{Fig3HeEv}, 
   lower panel, the behaviour of $^3$He/H across the Galactic disc as 
   predicted by the same models for the present epoch is displayed and 
   compared to H\,{\small II} region observations by Bania et al. (2002). We 
   do not recover the flat behaviour suggested by Bania et al. (2002), but 
   rather predict a negative $^3$He gradient in the 4\,--\,18 kpc 
   Galactocentric distance range, unless the case of 100 per cent of low-mass 
   stars suffering extra-mixing is considered.

   As far as $^4$He is concerned, starting from the {\it WMAP\/} value of 
   $Y_{\mathrm{p}} = 0.248$, a value of $Y_\odot = 0.261$ (Model CRM) and 
   $Y_\odot = 0.27$ (Model Tosi 1) is found at the time of Sun formation. 
   The latter nicely compares to the value at birth of the Sun of $Y = 0.275 
   \pm 0.01$ from Grevesse \& Sauval (1998; see also Bahcall, Pinsonneault, \& 
   Basu 2001). Notice that by adopting the Tinsley (1980) IMF, i.e. the same 
   IMF assumed by Model Tosi 1, Model CRM predicts $Y_\odot = 0.274$. A 
   detailed discussion of the effects of changing the IMF in the chemical 
   evolution code will be presented in a forthcoming paper.

%
   \begin{figure}
   \centerline{\psfig{figure=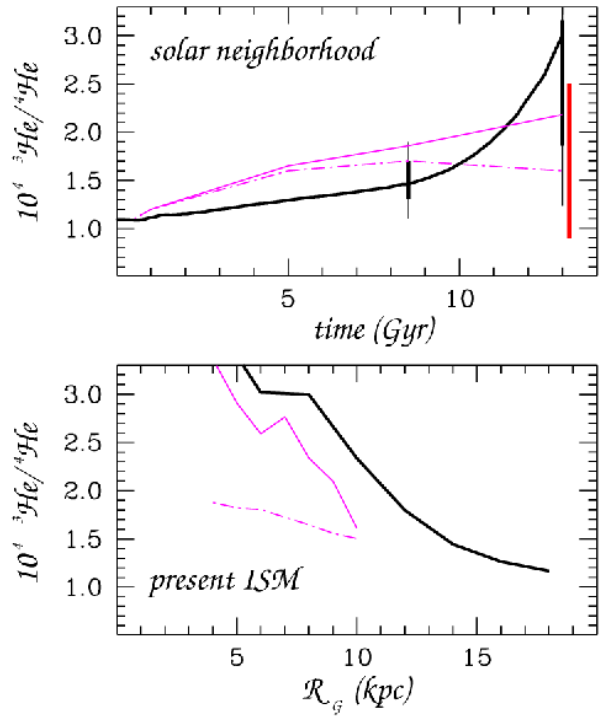,width=0.48\textwidth,angle=0}}
      \caption{Temporal (upper panel) and spatial (lower panel) variation of 
	      the helium isotopic ratio in the solar neighbourhood and across 
	      the Galactic disc at the present time, respectively. The 
	      predictions from Models CRM (thick solid lines) and Tosi 1 with 
	      93 and 100 per cent of extra-mixing (thin solid and dot-dashed 
	      lines, respectively) are compared to the observations (see text 
	      for details).
              }
         \label{Fig3He4He}
   \end{figure}
%

%
   \begin{figure*}
   \centerline{\psfig{figure=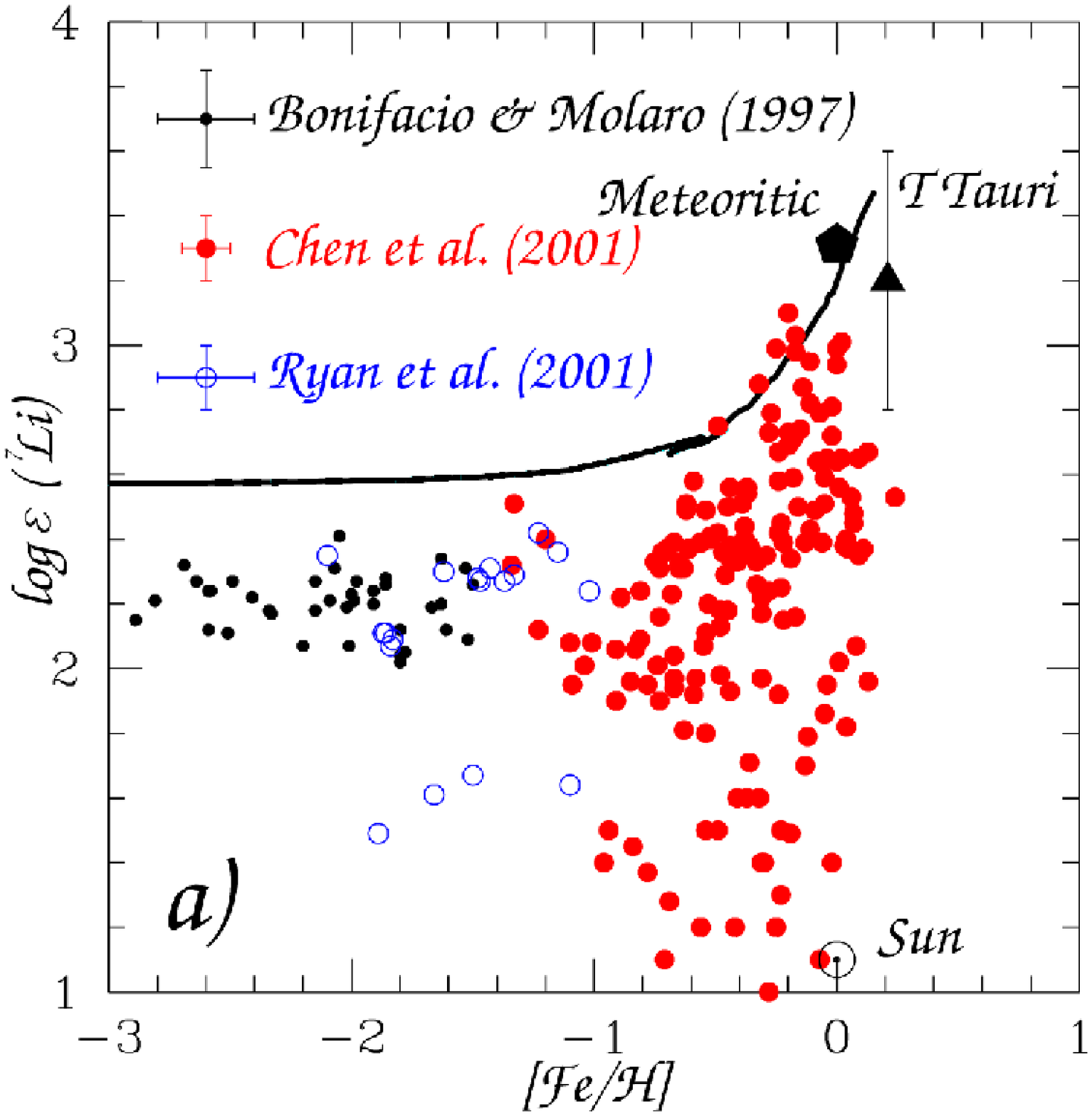,width=0.48\textwidth,angle=0}
               \psfig{figure=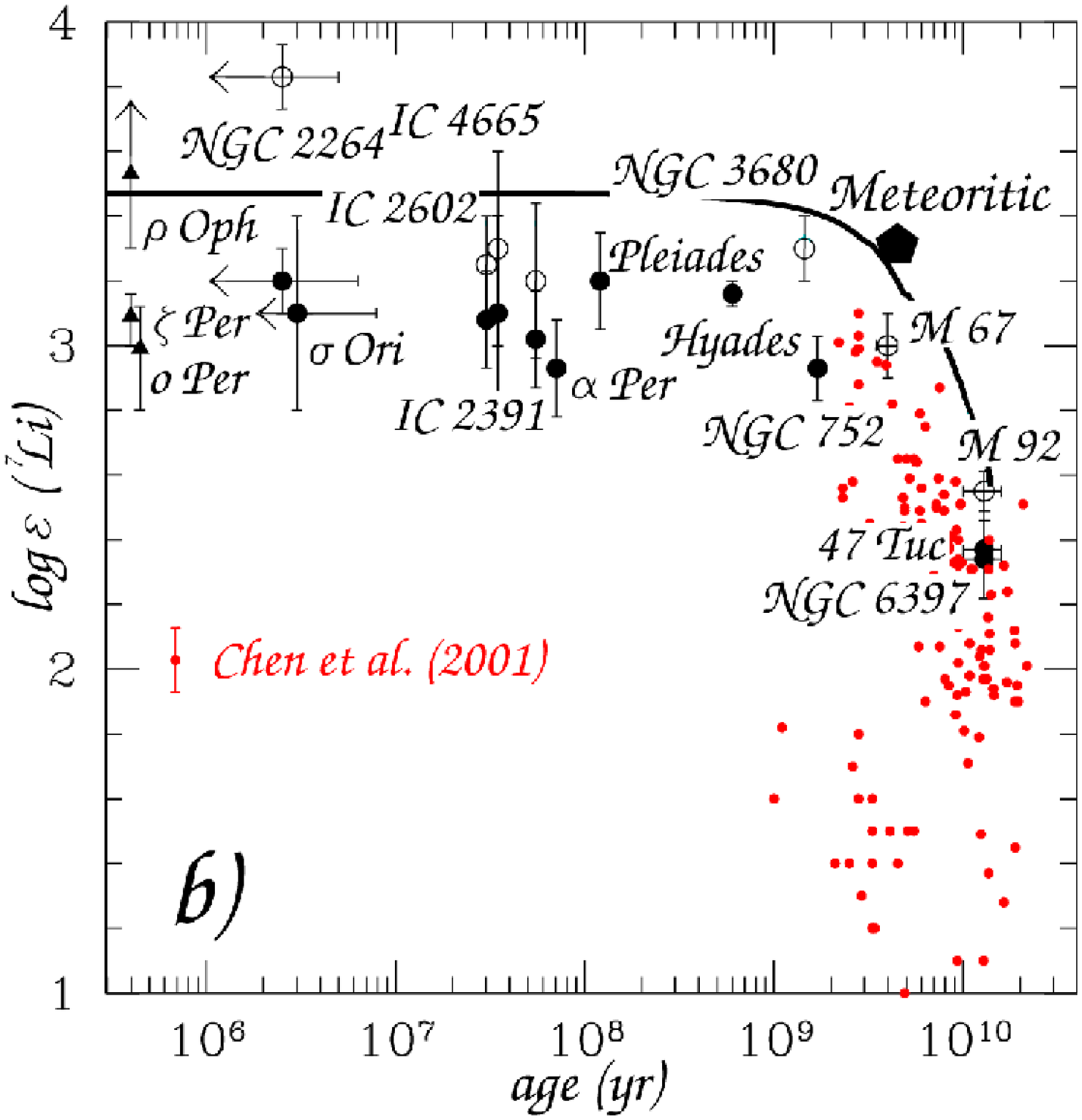,width=0.48\textwidth,angle=0}}
      \caption{a) Evolution of $^7$Li in the solar neighbourhood. The 
	      continuous line is a model adopting the prescriptions of the 
	      best model of Romano et al. (2001), except for the fact that it 
	      starts from a higher primordial lithium abundance of $\log 
              \varepsilon$($^7$Li)$_{\mathrm{p}}$ $\sim$ 2.6. Data are non-LTE 
	      values for dwarf stars hotter than 5700 K from Bonifacio \& 
	      Molaro (1997), Chen et al. (2001) and Ryan et al. (2001). The 
	      solar (Grevesse \& Sauval 1998), meteoritic (Nichiporuk \& Moore 
	      1974) and T\,Tauri star (at 2$\sigma$ error; Stout-Batalha et 
	      al. 2000) abundances are also shown with different symbols. b) 
	      Evolution of $^7$Li in the solar neighbourhood in a $\log 
	      \varepsilon$($^7$Li) versus age diagram. The chemical evolution 
	      model is the same as in Fig.~5a. Data for Galactic open clusters 
	      and globular clusters (big circles) are shown with their error 
	      bars as well as different measurements of the ISM value 
	      (triangles) (see text for references). The meteoritic value 
	      (pentagon) and data for field dwarfs with [Fe/H] $\ge -1.4$ 
	      (small filled circles) are also shown.
              }
         \label{FigLiEv1}
   \end{figure*}
%

   In Fig.~\ref{Fig3He4He} we show the temporal (upper panel) and spatial 
   (lower panel) variation of the helium isotopic ratio in the solar 
   neighbourhood and across the Galactic disc at the present time, 
   respectively. Only the predictions from Models CRM and Tosi 1 with 93 per 
   cent and 100 per cent of extra-mixing are shown. The data for the solar 
   nebula and the local ISM (vertical bars in the upper panel) are taken from 
   Geiss \& Gloeckler (1998; at 1 and 2$\sigma$ -- thick and thin lines, 
   respectively) and Salerno et al. (2003; 1$\sigma$-bar on the right at $t = 
   13$ Gyr). The measurement of $^3$He/$^4$He in the local ISM from the 
   COLLISA experiment on board of {\it MIR\/} [$^3$He/$^4$He = (1.7 $\pm$ 0.8) 
   $\times$ 10$^{-4}$; Salerno et al. 2003] involves neutral gas reaching 
   Earth's orbit before interacting with the solar EUV photons, which thus 
   keeps the original isotopic abundance ratios. If heavier weight is given to 
   this more recent measurement, a case where {\emph{nearly all}} low-mass 
   stars suffer extra-mixing should be preferred. A reliable estimate of the 
   $^3$He/$^4$He gradient across the Galactic disc is hence highly desirable 
   in order to establish which is the actual fraction of low-mass stars 
   destroying their pristine $^3$He. In fact, different models, assuming 
   different degrees of extra-mixing, predict large differences in the 
   $^3$He/$^4$He ratio across the Galactic disc, especially at inner radii 
   (Fig.~\ref{Fig3He4He}, lower panel). Assessing this point could have 
   important consequences on our knowledge of the mechanisms of $^3$He 
   production/destruction inside stars. Unfortunately, the value of the 
   $^3$He/$^4$He ratio at different positions along the Galactic disc 
   determined by observations is very uncertain, owing to the uncertain, 
   model-dependent ionization corrections which apply to the raw data.

   The temporal variation of D/H and $^3$He/H in the solar neighbourhood 
   predicted by the two different chemical evolution models (Model CRM and 
   Model Tosi 1), when starting from the same primordial abundances and 
   adopting the same nucleosynthesis prescriptions, is fairly similar (see 
   Figs.~\ref{FigDeutEv} and \ref{Fig3HeEv}). However, Model Tosi 1 predicts a 
   slightly lower deuterium abundance and a slightly higher $^3$He abundance 
   from $t \approx 5$ Gyr to now, except for the last 1 Gyr of evolution, when 
   a higher $^3$He mass fraction in the ISM is expected according to Model CRM 
   than according to Model Tosi 1. The same behaviour is recovered also in the 
   $^3$He/$^4$He versus time plot (Fig.~\ref{Fig3He4He}). The differences in 
   the abundance trends can be explained as due to the different stellar 
   lifetimes, IMF slopes, infall and star formation laws adopted by the two 
   models. In particular, in Model Tosi 1 stars in the mass range 0.5\,--\,0.9 
   $M_\odot$ never die, while they do in Model CRM, and these low-mass stars 
   contribute an important fraction of $^3$He at late times. The different 
   choice of the IMF produces a smaller -- although still non-negligible -- 
   effect on the predicted $^3$He evolution.

   \subsection{Evolution of $^{\mathbf{7}}$Li in the solar neighbourhood}

%
   \begin{figure*}
   \centerline{\psfig{figure=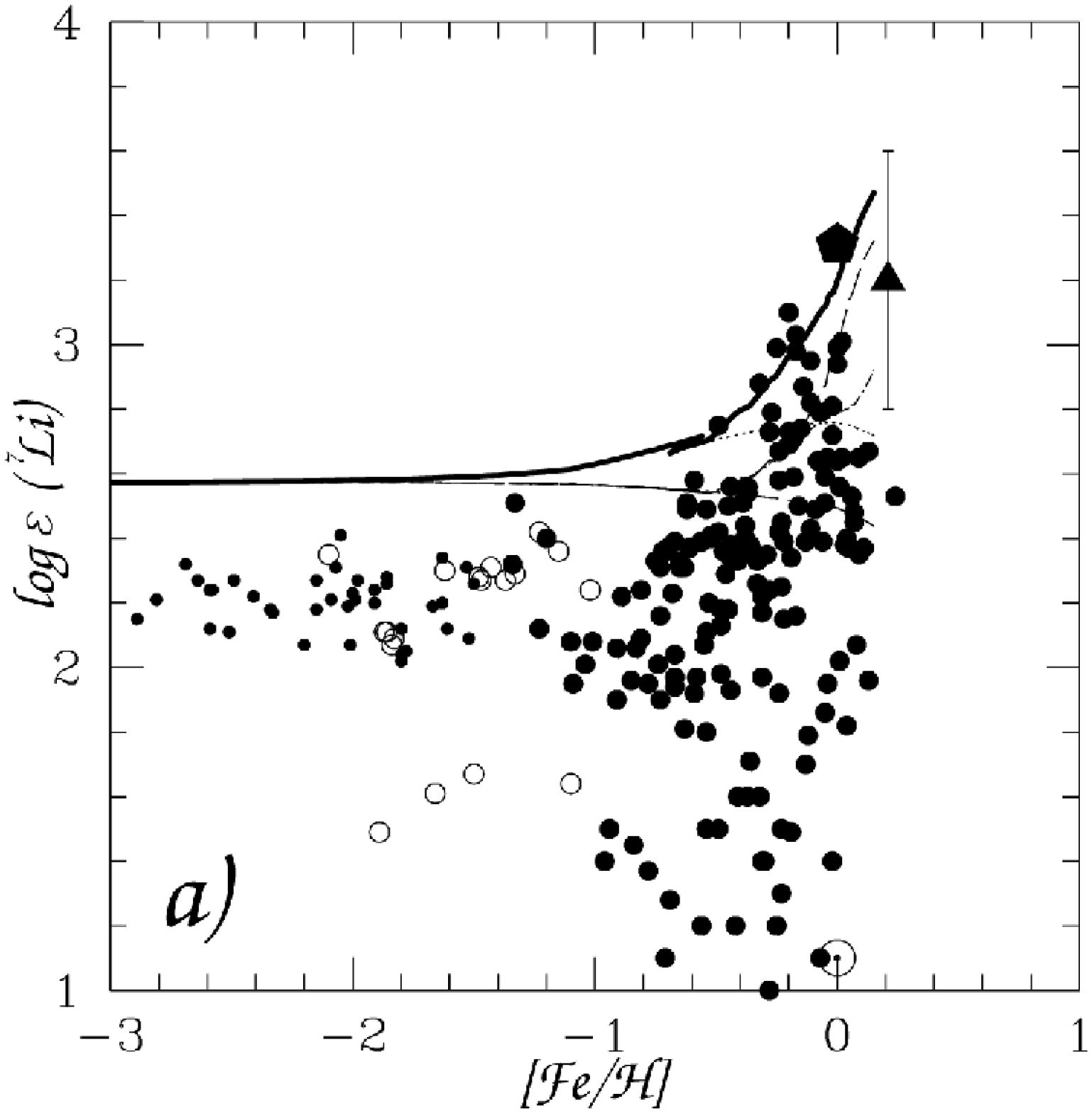,width=0.48\textwidth,angle=0}
               \psfig{figure=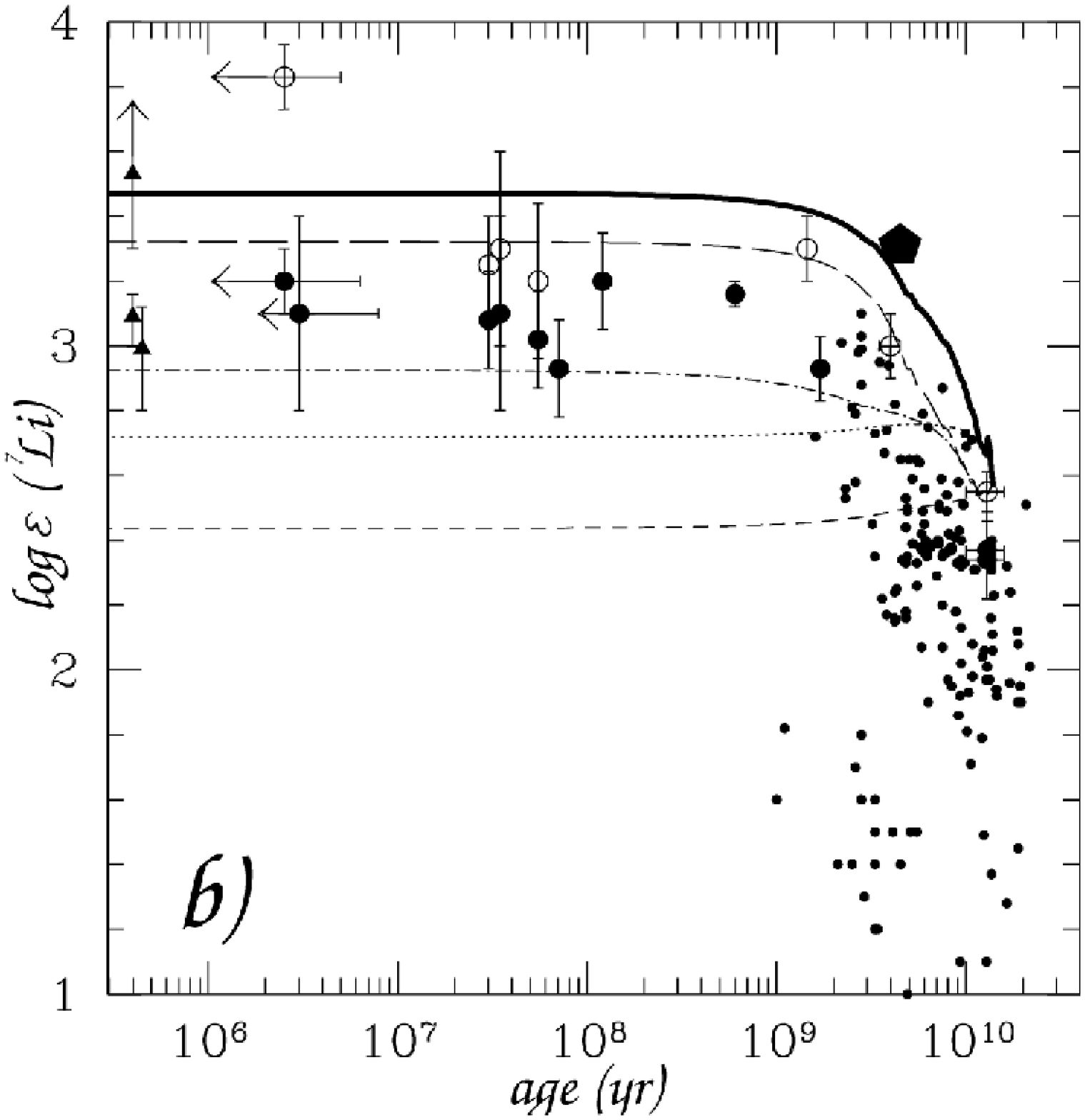,width=0.48\textwidth,angle=0}}
      \caption{Same as Fig.~\ref{FigLiEv1}, but here we show also the single 
	      contributions to the Li enrichment by the different stellar 
	      lithium factories we consider: AGB stars (short-dashed line); 
	      SNeII (dotted line); low-mass red giants (long-dashed line); 
	      novae (dot-dashed line).
              }
         \label{FigLiEv2}
   \end{figure*}
%

   Once the SBBN model predictions from Fiorentini et al. (1998) are adopted, 
   the {\it WMAP\/} $\eta_{\mathrm{10}}$ value of 6.1 results in a primordial 
   lithium abundance of ($^7$Li/H)$_{\mathrm{p}} \approx 3.75 \times 10^{-10}$ 
   (see Fig.~\ref{FigLightPrim}). In the usual $\log \varepsilon$($^7$Li) 
   notation, this translates to $\log \varepsilon$($^7$Li)$_{\mathrm{p}} \sim 
   2.6$, higher than any previous estimate of the primordial $^7$Li abundance 
   from observations of warm, metal-poor halo dwarfs under the hypothesis that 
   they have preserved all their pristine lithium and have not been 
   significantly polluted by early Galactic lithium production. In the 
   following, we analyse the effect of adopting such a high value for the 
   primordial $^7$Li abundance in the chemical evolution code.

   In Figs.~5a and 6a, we show the $\log \varepsilon$($^7$Li) versus [Fe/H] 
   diagram for field dwarfs in the solar neighbourhood hotter than 5700 K 
   (data are from Bonifacio \& Molaro 1997; Ryan et al. 2001; Chen et al. 
   2001). The solar (Grevesse \& Sauval 1998), meteoritic (Nichiporuk \& Moore 
   1974) and T\,Tauri (Stout-Batalha, Batalha \& Basri 2000) values are shown 
   as well. In the context of standard stellar evolution, warm halo stars 
   ([Fe/H] $\le -$1.5 dex) should not display lithium abundances significantly 
   lower than that in the gas out of which they formed (Deliyannis, Demarque 
   \& Kawaler 1990). In fact, in standard stellar evolution models lithium is 
   destroyed mainly during the pre-main sequence phase (during the main 
   sequence lifetime, the convective regions are too shallow to allow Li 
   depletion) and pre-main sequence Li depletion is not expected at low 
   metallicities (D'Antona \& Mazzitelli 1984). However, the halo stars could 
   destroy a significant fraction of their initial lithium content, if they 
   were suffering for non-standard destruction mechanisms (e.g., Vauclair \& 
   Charbonnel 1995, 1998; Th\'eado \& Vauclair 2001; Pinsonneault et al. 1999, 
   2002). Since these non-standard depletion mechanisms are responsible for a 
   broad spread in the Li abundances, as actually seen in the data at [Fe/H] 
   $\ge -$1.0 dex, the remarkable flatness and thinness of the \emph{Spite 
   plateau} represent a major challenge to stellar evolution models taking 
   non-standard $^7$Li depletion into account. Recent models considering 
   gravitational waves as the major depletion cause (Talon \& Charbonnel 2003) 
   seem indeed capable of reproducing both the observed $^7$Li spread at high 
   metallicities and the tightness of the \emph{Spite plateau} at low ones. 
   These models predict that the observed \emph{Spite plateau} is actually 
   4\,--\,5 times lower than the primordial value (Charbonnel 2003, private 
   communication).

   In Figs.~5b and 6b, we show the $\log \varepsilon$($^7$Li) versus age 
   diagram, which allows for a better appreciation of the lithium variations 
   in the last 1 Gyr. Data for Galactic open clusters as well as old globular 
   clusters are shown. The big filled circles represent mean lithium values, 
   while the big open circles represent the highest lithium abundance observed 
   in the cluster. Different measurements of the ISM value are also shown. For 
   NGC\,6397, the mean value for 12 turnoff stars corrected for non-LTE and 
   standard depletion is that by Bonifacio et al. (2002). For 47\,Tuc, the 
   $^7$Li abundance is the mean value for 2 turnoff stars determined by 
   Pasquini \& Molaro (1997). For M\,92, the highest non-LTE Li abundance 
   measured by Boesgaard et al. (1998) is shown. The age of all these clusters 
   is taken from Carretta et al. (2000). For M\,67, data are from Pasquini, 
   Randich \& Pallavicini (1997); for NGC\,752, from Balachandran (1995 -- 
   average of the three stars with the largest Li abundances on the hot side 
   of the Li dip). For NGC\,3680, we assume the highest lithium abundance from 
   Pasquini, Randich \& Pallavicini (2001); for the Hyades, the mean value is 
   taken from Balachandran (1995 -- average of the three stars with the 
   largest Li abundances on the cool side of the Li dip). The values for the 
   Pleiades and $\alpha$ Per are from Soderblom et al. (1993) and Randich et 
   al. (1998), respectively, while those for IC\,2391 and IC\,2602 are non-LTE 
   values from Randich et al. (2001). For IC\,4665 and NGC\,2264, non-LTE 
   values from Mart\'\i n \& Montes (1997) and Soderblom et al. (1999), 
   respectively, are displayed. The value for $\sigma$ Ori is the mean value 
   from Zapatero Osorio et al. (2002). The ISM values are from Lemoine et al. 
   (1993; line of sight towards $\rho$ Oph) and Knauth et al. (2000; lines of 
   sight towards $o$ Per and $\zeta$ Per). Data for field dwarfs (Chen et 
   al. 2001) are also shown for comparison.

   The thick continuous lines in Figs.~\ref{FigLiEv1} and \ref{FigLiEv2} 
   represent the predictions of a model where Galactic cosmic ray (GCR) and 
   stellar lithium production are taken into account in order to rise the 
   $^7$Li content in the ISM from its primordial value of $\log 
   \varepsilon$($^7$Li)$_{\mathrm{p}} \sim 2.6$ to the meteoritic [$\log 
   \varepsilon$($^7$Li) = 3.3] and local [$\log \varepsilon$($^7$Li) $\sim$ 
   3.2] ones. In particular, lithium is mostly produced by low-mass, 
   long-lived stellar sources (low-mass stars on the red giant branch and 
   novae). Only a minor contribution comes from stars on the asymptotic giant 
   branch and from Type II supernovae (see Romano et al. 1999, 2001 and 
   Fig.~\ref{FigLiEv2}) contrary to what has been suggested by Travaglio et 
   al. (2001), who invoke a large contribution from AGB stars (see, however, 
   Ventura, D'Antona \& Mazzitelli 2002 for a critical analysis of their 
   results). It is found that the amount of stellar production required in 
   order to match the meteoritic data does not change with changing the 
   assumed primordial $^7$Li abundance from $\log 
   \varepsilon$($^7$Li)$_{\mathrm{p}} \sim 2.2$ to $\log 
   \varepsilon$($^7$Li)$_{\mathrm{p}} \sim 2.6$, owing to the fact that the 
   evolution of $^7$Li during almost the whole Galaxy evolution is practically 
   determined solely by the amount of lithium produced through stellar and GCR 
   processes (cf. figure 6 of Romano et al. 2001). The different lines in 
   Fig.~\ref{FigLiEv2} represent the predictions of models where lithium is 
   contributed by only a single stellar Li factory: asymptotic giant branch 
   (AGB) stars (short-dashed line); Type II supernovae (SNeII; dotted line); 
   low-mass red giants (long-dashed line); novae (dot-dashed line).

   We conclude that, even in the case in which the primordial abundance of 
   lithium is as high as $\log \varepsilon$($^7$Li)$_{\mathrm{p}} \sim 2.6$ 
   (as suggested from recent {\it WMAP\/} data), our previous conclusions on 
   the $^7$Li evolution in the solar neighbourhood are left unchanged; in 
   particular, the rise off the primordial plateau value is still explained as 
   due to the same important, late $^7$Li contribution from long-lived stellar 
   sources (low-mass red giants and novae; Romano et al. 1999, 2001). However, 
   if the $\log \varepsilon$($^7$Li)$_{\mathrm{p}}$ value is as high as $\sim 
   2.6$, we strongly need some non-standard depletion mechanism able to 
   explain the absence of scatter and the high level of flatness observed in 
   halo stars over a quite large range of metallicity.


   \section{Final remarks and conclusions}

   The recent results from {\it WMAP\/} and their direct consequences on the 
   primordial abundances of the light elements have finally allowed us to 
   check that GCE models able to reproduce all the major observed properties 
   of the Milky Way are consistent also with SBBN predictions. It has been 
   shown several times in the last decade (e.g., Steigman \& Tosi 1992; Galli 
   et al. 1995; Prantzos 1996; Tosi 1996; Tosi et al. 1998; Chiappini et al. 
   2002) that only a very moderate D depletion from its primordial abundance 
   to the present one is allowed to let the models reproduce the observed 
   radial distributions of chemical abundances, star and gas densities and 
   star formation rates, as well as the age-metallicity relation and the 
   G-dwarf metallicity distribution. However, until now there was no 
   definitive observational evidence on the primordial values, since one 
   cannot completely exclude that even high-redshift, low-metallicity QSO 
   absorbers might be already polluted by stellar nucleosynthesis. Now we have 
   seen that the D, $^3$He and $^4$He produced during the big bang are in 
   excellent agreement with Galactic evolution requirements. For $^7$Li, 
   some problems do arise, but they mainly concern our understanding of the 
   mechanisms of lithium dilution/destruction in stars rather than chemical 
   evolution. Indeed, by assuming the {\it WMAP\/} primordial abundance of 
   $\log \varepsilon$($^7$Li)$_{\mathrm{p}} \sim 2.6$, we still need the same 
   important, late contribution of $^7$Li from long-lived stellar sources 
   (low-mass red giants and novae) required in order to explain the 
   observations in meteorites when starting from the lower value of $\log 
   \varepsilon$($^7$Li)$_{\mathrm{p}} \sim 2.2$ suggested from observations of 
   halo stars under the hypothesis that they neither destroyed their pristine 
   lithium nor suffered any pollution by early GCE (see also Romano et al. 
   2001). In fact, the amount of lithium production from stars and GCRs 
   required by our model in order to explain the observations is such that it 
   largely overwelms the lithium primordial abundance. However, if the adopted 
   primordial value of $\log \varepsilon$($^7$Li)$_{\mathrm{p}} \sim 2.6$ is 
   confirmed, a mechanism able to deplete Li in halo stars while preserving 
   the flatness of the \emph{Spite plateau} and producing almost no scatter in 
   the star to star abundance at low metallicities is strongly needed. 
   Gravitational waves seem interesting in this respect (see Talon \& 
   Charbonnel 2003), but we still wait for full metallicity dependent 
   computations.

   We should recall at this point that our standard chemical evolution models 
   refer to large-scale, long-term phenomena and cannot account for 
   small-scale, short-term variations. If we want to reproduce also the 
   observed spread in the abundances of the light elements, as observed, for 
   instance, by {\it FUSE\/} in the local ISM (e.g., Moos et al. 2002), or for 
   $^3$He/H across the Galactic disc (Bania et al. 2002), we should also take 
   into account the possible inhomogeneities in the chemical enrichment of 
   each region and the effects of possible orbital diffusion of the stars.


\section*{Acknowledgments}

      M.T. is particularly grateful to Corinne Charbonnel, Johannes Geiss, and 
      George Gloeckler of the LOLA-GE team for the enlightening discussions at 
      the International Space Science Institute in Berne (CH). Dana Balser, 
      Tom Bania, and Bob Rood are warmly thanked for always being ready to 
      share updated values of the $^3$He abundances. We also thank Daniele 
      Galli and Gary Steigman for their useful comments. Ed Jenkins, Warren 
      Moos and Alfred Vidal-Madjar are gratefully acknowledged for clarifying 
      what is the range of LISM D/H values which should be quoted according to 
      the most reliable data. This work has been partially supported by the 
      {\emph{Italian ASI}} through grant IR\,11301ZAM.

\bsp

\label{lastpage}

\end{document}